\input{aipcheck}

\documentclass[
    ,final            
,sort&compress]{aipproc}

\layoutstyle{8x11single}
\usepackage{amsfonts}

\begin{document}

\title{Statistical scattering of waves in disordered waveguides: The limiting macroscopic statistics in the ballistic regime}



\author{M. Y\'epez}{address={Departamento de F\'isica de la Materia Condensada, Universidad Aut\'onoma de Madrid, E-28049 Madrid, Spain}}

\author{P. A. Mello}{address={Instituto de F\'isica, Universidad Nacional Aut\'onoma de M\'exico, 01000 M\'exico, Distrito Federal, M\'exico}}

\author{J. J. S\'aenz}{address={Departamento de F\'isica de la Materia Condensada, Universidad Aut\'onoma de Madrid, E-28049 Madrid, Spain}}

\begin{abstract}

In this work, we present a theoretical study of the statistical properties of wave scattering in a disordered ballistic waveguide of length $L$; we have called this system the ``building block''. The building block is interesting as a physical system because its statistical properties could be studied experimentally in the laboratory.

In order to study the building block, as a physical system in itself, we have developed a perturbative method based on Born series. This method is valid only in the ballistic regime, when the length of the system $L$ is smaller than the mean free path $\ell$, and in the short-wave-length approximation, when the the wave number $k$ and the mean free path $\ell$ satisfy $k\ell \gg 1$. This method has allowed to find, analytically, the behavior of quantities of interest that we have not been able to find from the diffusion equation. In contrast with the diffusion equation method, which takes into account approximately the contribution of closed channels, this method takes them explicitly. In earlier works numerical evidence was found that the expectation values of some interesting quantities are insensitive to the number of closed channels that were used on the calculations; with this method, we could show that closed channels are relevant for the expectation values of amplitudes but irrelevant for the intensities and conductance expectation values. The results of this method show a good agreement with numerical simulations.

\end{abstract}

\maketitle

\section{Introduction}

The statistical of certain \emph{complex} wave interference phenomena, such as the fluctuations of transmission and reflection of waves, is of considerable interest in many fields of physics \cite{AIshimaru:1978,BLAltshuler:1991,PSheng:1995}. The complexity derives from the randomness of the scattering potentials, as in the case of disordered conductors with impurities, or more in general, disordered waveguides. It is the latter domain that will interest us here. The complexity no only derives from the randomness of the potential, it is also consequence of the multiple scattering processes. The study will focus in the context of quantum mechanics, so we will talk about electron or scalar waves, but the method could be extended to classical waves; electromagnetic or elastic waves.

Due to the randomness of the potential, a statistical approach is needed because any change in the microscopic realization of the disorder will completely change the interference pattern \cite{Stone2008}; therefore, only a statistical description makes sense, so the aim is to obtain the expectation values of macroscopic observables related with the transport properties as function of the length $L$ of the system. The expectation values will be computed as an averaged over an ensemble and we will denote them by $\left\langle \cdots\right\rangle_{L}$; for instance $\left\langle g\right\rangle_{L}$ denotes the expectation value of the conductance.

Previous studies have been focused in the diffusive and localized regimes (see figure \ref{Theregimes}), where remarkable statistical regularities have been found in the sense that the probability distribution for various macroscopic quantities involves a rather small number of physical parameters; the mean free path $\ell$. These statistical regularities showed the existence of a \emph{limiting macroscopic statistics} for those regimes. In the diffusive regime and in the bulk disorder case, the DMPK equation (Dorokhov \cite{Dorokhov:1982} and Mello et al. \cite{AnnPhys181:1988}) describes successfully the statistical properties of the conductance \cite{PRB37:1988, AnnPhys181:1988}. 

The motivation of this work is to study the statistical properties of wave scattering in the ballistic regime ($L \ll \ell$ see figure \ref{Theregimes}), which has not been as studied as the diffusive and localized regimes. In a more recent work \cite{PRE75:2007} we have obtained a more general diffusion equation than the DMPK equation that gives a more satisfactory physical description than DMPK equation. In the derivation of that new equation, the statistical properties of a ballistic system, call building block, played an essential role. In this work es shall use a perturbative method based on the Born series to shown that in the ballistic regime also exists a limiting macroscopic statistics.

\section{The microscopical model of the potential and the scattering problem.}

\begin{figure}[h]
\includegraphics[height=0.12\textheight]{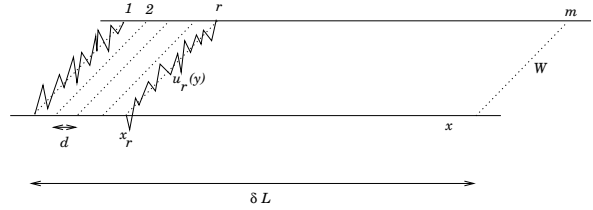}
\caption{Construction of the building block using $\delta$-potential slices.}
\label{deltasecuence}
\end{figure}

The building block is constructed as a sequence of $m\gg 1$ equidistant scattering units in the longitudinal (or wave propagation) direction $x$, separated by distance $d$, so the length of the system is fixed and given by $L = m d$ . The potential strength of the scatters obeys a statistical distribution that will be explained below. Each scattering unit is assumed to be a $\delta$-potential ``slice'' in the longitudinal direction $x$ and to have a random behavior $u_{r}\left(y\right)$ in the transversal direction $y$ as shown schematically in Fig. \ref{deltasecuence}; therefore, the total potential of the building bock is given by the following expression

\begin{equation}
U\left( x,y \right)=\sum_{r=1}^{m}U_{r}\left(x,y\right)=\sum_{r=1}^{m}u_{r}\left(y\right)\delta\left(x-x_{r}\right), \label{Mic_Pot_Mod}
\end{equation}

\noindent where the $r$th $\delta$ slide potential is defined as $U_{r}\left(x,y\right)=u_{r}\left(y\right)\delta\left(x-x_{r}\right)$, being $u_{r}\left(y\right)$ a random function with an arbitrary dependence in the $y$ coordinate.

\noindent The physical regime in which we shall work is such that the separation $d$ between slices is much smaller than the wavelength $\lambda$ of the incident wave, the length $L$ of the building block and the mean free path $\ell$, i.e.,

\begin{equation}
d \ll \left\lbrace \lambda, L, \ell\right\rbrace;
\end{equation}

\noindent the relations between $\lambda$, $L$ and $\ell$ will be specified later.

Since the Schr\"odinger has to be solved with Dirichlet boundary conditions at the walls of the waveguide, a discretization in the transversal direction appears; that discretization is given by the transverse states

\begin{equation}
\chi_{b}\left( y\right) = \sqrt{\frac{2}{W}}\sin\frac{\pi by}{W},\label{Tran_Mods}
\end{equation}

\noindent which vanish at the walls, $y=0$, $W$, if the ``channel'' or ``mode'' index $b$ is an integer number. We must note that if $N\pi < kW <\left(N+1\right)\pi$ the waveguide admits precisely $N$ open channels; therefore, if $1\leq b\leq N $ we are taking about of an open channel or traveling mode, while if $b > N$ we refer to $b$ as a closed channel or evanescent mode.

\begin{figure}[h]
\includegraphics[scale=0.17]{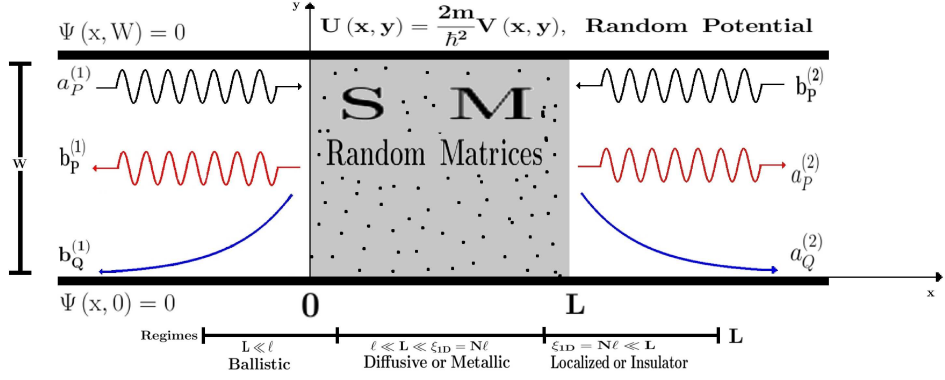}
\caption{Scattering problem and physical regimes. The quantities $a_{P}^{\left(1\right)}$, $b_{P}^{\left(2\right)}$ are $N$ dimensional vectors, being its elements all the possible incoming open channel amplitudes; analogously, $b_{P}^{\left(1\right)}$, $a_{P}^{\left(2\right)}$ are vector with all the possible outgoing open channel amplitudes and $b_{Q}^{\left(1\right)}$, $a_{Q}^{\left(2\right)}$, are the corresponding outgoing closed channel vectors.}
\label{Theregimes}
\end{figure}

In order to describe the scattering problem, we consider that an incoming wave (particle) in the open channel $a_{0}$ is traveling from left to right and scattered by the potential \eqref{Mic_Pot_Mod}, giving as result, outgoing waves both in open channels as in closed channels. Due to our interest is to give a perturbative approach to the scattering problem, we will describe the problem in terms of the scattering matrix $S$, which relates the outgoing wave amplitudes of open channels (wavy outgoing lines in the figure \ref{Theregimes}) with the corresponding incoming wave amplitudes\footnote{The outgoing closed channels amplitudes $b_{Q}^{\left( 1\right)}$ and $a_{Q}^{\left( 2\right)}$, which decrease exponentially at both sides of the system, can be expressed in terms of incoming wave amplitudes, through the extended scattering matrix; see references \cite{MelloKumar:2004,PRE75:2007,PhDTMyepez:2009}} (wavy incoming lines in figure \ref{Theregimes}), i.e,

\begin{equation}
\left(
\begin{array}{c}
b_{P}^{\left(1\right)}
\\
a_{P}^{\left(2\right)}
\end{array}
\right)=S
\left(
\begin{array}{c}
a_{P}^{\left(1\right)}
\\
b_{P}^{\left(2\right)}
\end{array}
\right),
\;\;\;\;\;S=
\left(
\begin{array}{cc}
r & t^{\prime}
\\
t & r^{\prime}
\end{array}
\right). \label{ScatMat_Relat}
\end{equation}

\noindent Although the Eq. \eqref{ScatMat_Relat} gives a relation between open channels amplitudes, due to the multiple scattering processes inside the disordered system, the $S$ matrix has information related with scattering processes of closed channels which, as we shall show later, are very important for the statistics of the amplitudes, but no so for the statistics of the intensities.

The fundamental quantities we need studying are the probabilities amplitudes of transmission $t_{aa_{0}}$ and reflection $r_{aa_{0}}$, when an incident wave in the open channel $a_{0}$ is transmitted or reflected, respectively, in the open channel $a$. This can be done, in a perturbative way, by using the coupled Lippmann-Schwinger equations \footnote{For the longitudinal wave functions; for details see references \cite{MelloKumar:2004,PhDTMyepez:2009}.} and the corresponding integral expressions for the amplitudes $t_{aa_{0}}$ and $r_{aa_{0}}$; this procedure allows to find the Born series expansion for these quantities, which are series in powers of the matrix elements of the quasi-one-dimensional (Q1D) potential $U_{ab}\left(x\right)$

\begin{equation}
U_{ab}\left(x\right)=\int_{0}^{W}\chi_{a}\left(y\right)U\left(x,y\right) \chi_{b}\left(y\right)dy=\sum_{r=1}^{m}\left(u_{r}\right)_{ab}\delta\left(x-x_{r} \right),
\end{equation}

\noindent where the random quantities $\left(u_{r}\right)_{ab}$ are defined by the following expression:

\begin{equation}
\left(u_{r}\right)_{ab}=\int_{0}^{W}\chi_{a}\left(y\right)u_{r}\left(y\right) \chi_{b}\left(y\right)dy; \label{Q1D_Pot}
\end{equation}

\noindent therefore, the statistics of the amplitudes $t_{aa_{0}}$, $r_{aa_{0}}$, their corresponding intensities $T_{aa_{0}}=\left\vert t_{aa_{0}} \right\vert^{2}$, $R_{aa_{0}}=\left\vert r_{aa_{0}} \right\vert^{2}$, and the conductance $g$, need a statistical model for the Q1D potential $U_{ab}\left(x\right) $ or more precisely for the random quantities $\left( u_{r}\right) _{ab}$ of Eq. \eqref{Q1D_Pot}.

\section{The statistical model}

The Born series expansion that we have exposed here is given, in a naturally way, in terms of the quantities $\left(u_{r}\right)_{ab}$; therefore, we specify the statistical model in terms of those quantities. The $m$ potential matrix elements $\left(u_{r}\right)_{ab}$ ($r=1,\ldots,m$) are assumed to be statistically independent, identically distributed, with zero average and, for simplicity, zero odd moments, so that, for example,

\begin{equation}
\left\langle \left(u_{r} \right) _{b_{1}b_{2}} \right\rangle =0,
\;\;\;\;\;\;\;\;\;\;\;\;
\left\langle \left(u_{r_{1}} \right) _{b_{1}b_{2}}\left(u_{r_{2}} \right) _{b_{1}b_{2}} \right\rangle =
\left\langle \left(u_{r_{1}} \right) _{b_{1}b_{2}} \right\rangle\delta_{r_{1}r_{2}},
\label{2moment}
\end{equation}

\noindent are the first and second moment, respectively. It is important to note that the channel indexes $b_{1}$, $b_{2}$, could be both open or closed channels. This allows to take into account explicitly the contribution of closed channels.

In order to calculate the expectation values related with the amplitudes $t_{aa_{0}}$ and $r_{aa_{0}}$, we define the so-called dense weak scattering limit (DWSL), in which the various scatterers are assumed to be very weak, their linear density $\nu=1/d$ very large, in such a way that the channel-dependent mean free paths $\ell_{b_{1}b_{2}}$, are fixed, so that:

\begin{eqnarray}
\frac{1}{\ell_{b_{1}b_{2}}}&=&\nu\left\langle \left\vert \left(r_{r}\right)_{b_{1}b_{2}}  \right\vert ^{2} \right\rangle \approx \frac{\left\langle \left(u_{r}\right)_{b_{1}b_{2}}^{2} \right\rangle }{4k_{b_{1}}k_{b_{2}}d},
\;\;\;1\leq b_{1},b_{2}\leq N,
\label{MFP_PP}
\\
\frac{1}{\ell_{b_{1}b_{2}}}&=&\nu\left\langle \left\vert \left(r_{r}\right)_{b_{1}b_{2}}  \right\vert ^{2} \right\rangle \approx \frac{\left\langle \left(u_{r}\right)_{b_{1}b_{2}}^{2} \right\rangle }{4k_{b_{1}}\kappa_{b_{2}}d},
\;\;\;1\leq b_{1}\leq N,\;\;\;b_{2}>N,
\label{MFP_QQ}
\end{eqnarray}

\noindent where $\left(r_{r}\right)_{b_{1}b_{2}}$ is the reflection amplitude of an individual delta scattering unit, $k_{b_{1}}$ is the open channel wave number and $\kappa_{b_{1}}$ is the closed channel attenuation factor. These channel-dependent mean free paths depend explicitly on the energy through the quantities $k_{b_{1}}$ and $\kappa_{b_{1}}$.

In the DWSL the most important result emerges that \emph{the dependence on the moments of the potential higher than the second drops out in this limit}; therefore, the expectation values depend only on the second moments of the potential, Eq. \eqref{2moment}, through the mean free paths, Eqs \eqref{MFP_PP}-\eqref{MFP_QQ}; higher-order moments do not contribute in the DWSL. This result shows the existence of a generalized central limit theorem (CLT).

\section{Expectation values in the ballistic regime}

If we make use of the statistical model, Eq. \eqref{2moment}, the resulting  Born series is a double expansion: in powers of $L/\ell$ an in powers of $1/k\ell$, where we have used $k$ (the total wave number of the Schr\"odinger equation) and $\ell$ (the transport mean free path) to denote, symbolically, any longitudinal wave number $k_{b_{1}}$ and any mean free path $\ell_{b_{1}b_{2}}$, respectively; therefore, the Born series expansion will be only valid, in general, in the \emph{ballistic regime} and in the so call short wave length approximation (SWLA) or weak disorder approximation, i.e.,

\begin{equation}
L/\ell \ll 1, \;\;\;\;\; k\ell \gg 1.
\end{equation}

In the dense weak scattering limit (DWSL), expectation value of the transmission amplitude is given by the following expression:

\begin{eqnarray}
\lim_{DWS}\left\langle t_{aa_{0}}\right\rangle _{k,L}&=& \delta
_{aa_{0}}\biggl\{ \biggr[1-\left( \frac{L}{\ell _{a}}\right)
+\frac{1}{2!}\biggl[ \left( \frac{L}{\ell _{a}}\right)
^{2}-\left( \frac{L}{\ell _{a}^{\prime
}}\right) ^{2}\biggr]+\cdots \biggr]\label{taa0_until_4th_order_SWLA}
\\
&+&i\biggl[\left( \frac{L}{\ell _{a}^{\prime }}\right)
-\left( \frac{L}{\ell _{a}}\right) \left( \frac{L}{\ell
_{a}^{\prime }}\right)+\cdots \biggr]\biggr\}+O\left( \frac{1}{k\ell }\right),\nonumber
\end{eqnarray}

\noindent where we have defined, respectively, the \emph{scattering mean free paths} for the open channel $a$ related with transitions to open and closed channels, respectively:

\begin{equation}
\frac{1}{\ell_{a}}=\sum_{b_{1}=1}^{N}\frac{1}{\ell_{b_{1}a}},
\;\;\;\;\;\;\;\;\; 
\frac{1}{\ell_{a}^{\prime}}=\sum_{b_{1}=N+1}^{\infty}\frac{1}{\ell_{b_{1}a}},
\label{P_Q_Scat_MFP}
\end{equation}

\noindent As we can see from Eq. \eqref{taa0_until_4th_order_SWLA}, $\left\langle t_{aa_{0}}\right\rangle _{k,L}$ has a contribution from closed channels through $\ell_{a}^{\prime}$, which depends strongly on the number of closed channels that in principle is infinity. This scattering mean free path of closed channels $\ell_{a}^{\prime}$ gives as a result an \emph{imaginary part for expectation value of the transmission amplitude} that is not zero, which differs from the well known result in the SWLA $\left\langle t_{aa_{0}}\right\rangle _{k,L}=\delta_{aa_{0}}e^{-L/\ell_{a}}$ \cite{PRB46:1992}. This exponential behavior can be obtained from the Born series if the idealization of the SWLA, $k\ell \rightarrow \infty$ is considered and the contributions of the closed channels are neglected. Only for this observable and in the SWLA, we were able to obtain the dominant contribution in powers of $1/k\ell$; in this case the dominant contribution is $1/\left( k\ell\right) ^{0}$.

\begin{figure}[h]
\includegraphics[scale=0.08]{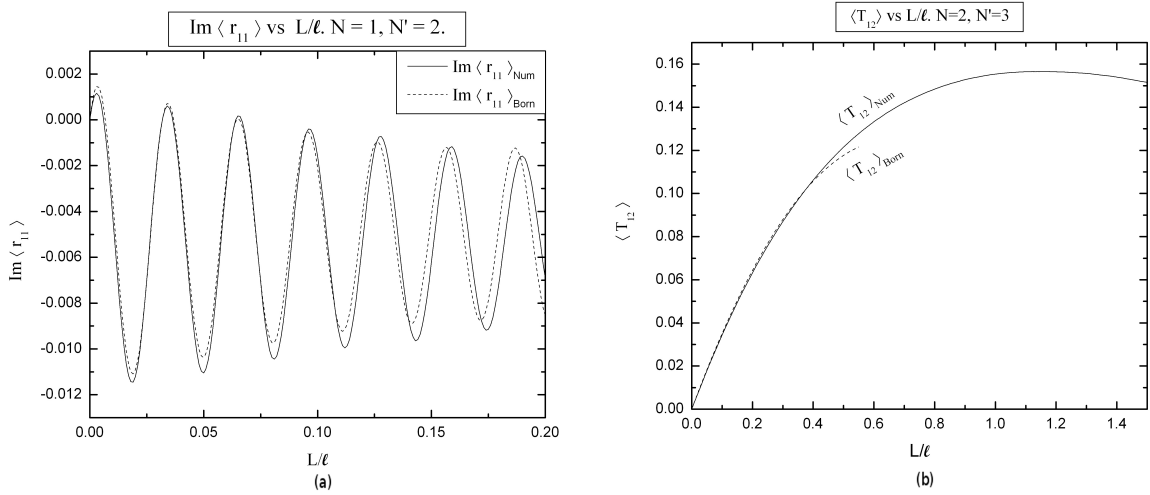}
\caption{Born series $vs$ numerical simulations. (a) $\mathrm{ Im}\left\langle r_{11}\right\rangle vs L/\ell$ when the waveguide supports $N=1$ open channel and we have taken into account $N^{\prime}=2$ closed channels. (b)  $\left\langle T_{12}\right\rangle vs L/\ell$ when the waveguide supports $N=2$ open channels and we have taken into account $N^{\prime}=3$ closed channels.}
\label{Imr11T12BornNum}
\end{figure}

The expectation value $\left\langle r_{aa_{0}}\right\rangle _{k,L}$ is obtained in an analogous way \cite{PhDTMyepez:2009}; once more, the Born series result differs from the well known result in the SWLA $\left\langle r_{aa_{0}}\right\rangle _{k,L}=0$ \cite{PRB46:1992}, and it depends explicitly on the closed channels. In figure \ref{Imr11T12BornNum}a we compare the Born series prediction for $\left\langle r_{11}\right\rangle _{k,L}$ with the corresponding result of a numerical simulation, when the waveguide supports $N=1$ open channels and we have taken into account $N^{\prime}=2$ closed channels. As we can see the agreement is very well in the ballistic regime.

In the dense weak scattering limit (DWSL), expectation value of the intensity $ T_{aa_{0}}=\left\vert t_{aa_{0}}\right\vert^{2}$ is given by the following expression:

\begin{eqnarray}
\lim_{DWS}\left\langle
T_{aa_{0}}\right\rangle_{k,L}&=&\delta_{aa_{0}}\left[1-2\frac{L}{\ell
_{a}}+2\left(\frac{L}{\ell_{a}}\right)^{2}+O\left(\frac{L}{\ell}\right)^{3}\right]\label{The_Taa0_kl_Expansion}
\\
&+&\left[\frac{L}{\ell
_{aa_{0}}}+\sum_{b_{1}=1}^{N}\frac{L^{2}}{\ell_{ab_{1}}\ell_{b_{1}a_{0}}}
-\frac{1}{\ell_{aa_{0}}}\left(\frac{1}{\ell_{a}}+\frac{1}{\ell_{a_{0}}}\right)L^{2}
+O\left(\frac{L}{\ell}\right)^{3}\right]+O\left(\frac{1}{k\ell}\right),\nonumber
\end{eqnarray}

\noindent where we can see that the dominant contribution of this quantity depends only on the open channels; the first contribution of the closed channels is of order $1/k\ell$. In figure \ref{Imr11T12BornNum}b we compare the Born series prediction for $\left\langle T_{12}\right\rangle _{k,L}$, Eq. \eqref{The_Taa0_kl_Expansion}, with the corresponding result of a numerical simulation, when the waveguide supports $N=2$ open channels and we have taken into account $N^{\prime}=3$ closed channels. As we can see the agreement is very well in the ballistic regime.

\begin{figure}[h]
\includegraphics[width=16.0cm, height=8.0cm]{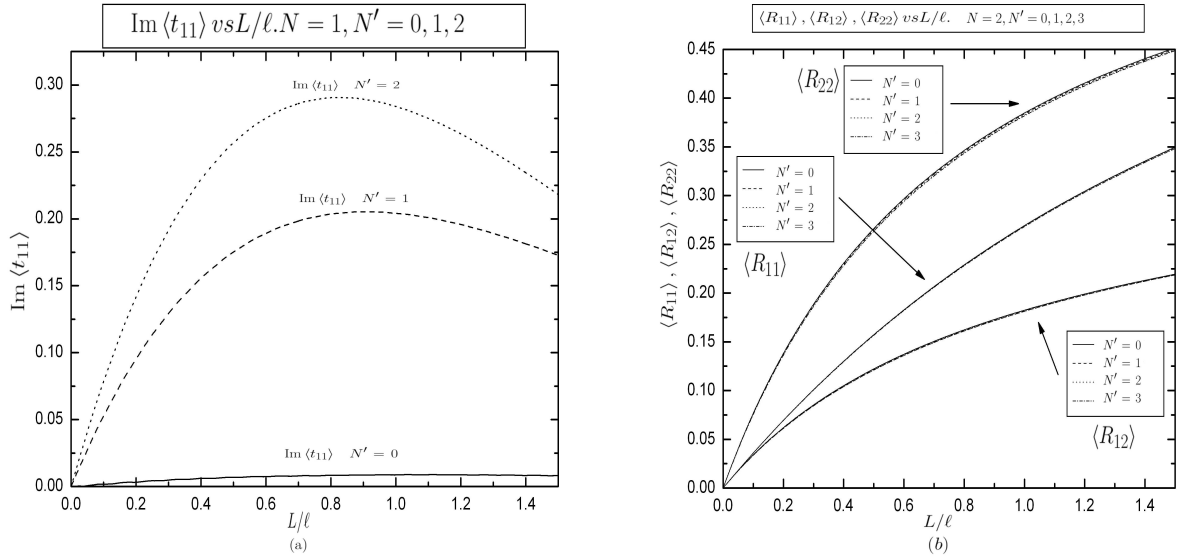}
\caption{Numerical evidence of the closed channels contribution beyond the ballistic regime. (a) $\mathrm{ Im}\left\langle t_{11}\right\rangle vs L/\ell$ when the waveguide supports $N=1$ open channel and we have taken into account $N^{\prime}=0,1,2$ closed channels. (b) $\left\langle R_{aa_{0}}\right\rangle vs L/\ell$ when the waveguide supports $N=2$ open channels and we have taken into account $N^{\prime}=0,1,2,3$ closed channels.}
\label{Imr11T12NumBeyodBR}
\end{figure}

The well agreement between Born series prediction and the numerical simulations, in the ballistic regime, shows that the contributions of closed channels are very important for the expectation values of amplitudes and irrelevant for the expectations values of intensities. The analytical dependence for amplitudes, Eq. \eqref{taa0_until_4th_order_SWLA}, and intensities, Eq. \eqref{The_Taa0_kl_Expansion}, suggest that the closed channel contributions could be important even beyond the ballistic regime; unfortunately, Born series method gives, in general, an \emph{asymptotic series} \cite{MorseyFeshbach} in powers of $L/\ell$, that approximate the expectation values only when $L/\ell \ll 1$, i.e., in the ballistic regime; therefore, we can not use this method beyond this regime; however, in figure \ref{Imr11T12NumBeyodBR}a, it is shown numerical evidence that, beyond the ballistic regime, $\mathrm{Im}\left\langle t_{11} \right\rangle $ ($N=1$ open channel) depends strongly on number of cosed channels that were used in the simulation; on the other hand, figure \ref{Imr11T12NumBeyodBR}b shows that, beyond the ballistic regime, $\left\langle R_{11} \right\rangle $, $\left\langle R_{12} \right\rangle $, $\left\langle R_{22} \right\rangle $ ($N=2$ open channels) do not depend, in an important way, on the number of closed channels that were used in the simulations.

\section{Conclusions}

We have developed a perturbative method (based on the Born series) to study the statistical properties of waves scattering in the ballistic regime $L/\ell \ll 1$ and in the short wave length approximation $k\ell \gg 1$. This method shows the existence of a \emph{limiting macroscopic statistics in the ballistic regime}, in the context that expectation values depend only on the microscopic details through the mean free paths $\ell_{ab}$, what represents a generalized \emph{central limit theorem}.

The Born series that we have obtained allows to obtain analytical information, in the ballistic regime, for quantities that we have not been able to obtain with the diffusion equation or Ref \cite{PRE75:2007}; this method takes into account, explicitly, the contributions of closed channels, which are essential to understand the statistics of amplitudes but those are irrelevant for the statistics of the intensities. 

The numerical results show a good agreement with the theoretical predictions, in the ballistic regime and exhibit that beyond the ballistic regime the contributions of closed are very important for the expectation values of amplitudes and irrelevant for those of the intensities.




\begin{theacknowledgments}

This work has been supported by the EU NMP3-SL-2008-214107-Nanomagma Project, the Spanish MICINN Consolider \textit{NanoLight} (CSD2007-00046) and FIS2009-13430-C02-02
and by the Comunidad de Madrid Microseres-CM (S2009/TIC-1476). M.Y also thanks Conacyt for its support through Grant No 179710 and Act. Carlos Lop\'ez Natar\'en from the computing technical secretariat at Instituto de F\'isica, UNAM, for his technical support.
  
\end{theacknowledgments}



\bibliographystyle{aipproc}   


\bibliography{references}

\IfFileExists{\jobname.bbl}{}
 {\typeout{}
  \typeout{******************************************}
  \typeout{** Please run "bibtex \jobname" to optain}
  \typeout{** the bibliography and then re-run LaTeX}
  \typeout{** twice to fix the references!}
  \typeout{******************************************}
  \typeout{}
 }

\end{document}